\documentclass[sigconf]{acmart}
\usepackage{algorithm2e}
\usepackage{balance}
\setcopyright{none}
\usepackage{balance}
\usepackage{caption}
\usepackage{tabularx, environ}
\usepackage{titlesec}
\usepackage{float}
\usepackage{ifsym}
\usepackage{hyperref}
\hypersetup{colorlinks,allcolors=black}

\copyrightyear{2020}
\acmYear{2020}
\setcopyright{acmcopyright}\acmConference[SIGIR '20]{Proceedings of the 43rd International ACM SIGIR Conference on Research and Development in Information Retrieval}{July 25--30, 2020}{Virtual Event, China}
\acmBooktitle{Proceedings of the 43rd International ACM SIGIR Conference on Research and Development in Information Retrieval (SIGIR '20), July 25--30, 2020, Virtual Event, China}
\acmPrice{15.00}

\begin{document}
\fancyhead{}

\title{JointMap: Joint Query Intent Understanding For Modeling Intent Hierarchies in E-commerce Search}

\author[1]{Ali Ahmadvand}
\authornote{Research was conducted while interning at The Home Depot Search \& NLP team.}
\affiliation{Emory University}
\email{ali.ahmadvand@emory.edu}

\author[2]{Surya Kallumadi}
\affiliation{\institution{The Home Depot}}
\email{surya@ksu.edu}

\author[3]{Faizan Javed}
\affiliation{\institution{The Home Depot}}
\email{faizan_javed@homedepot.com}

\author[4]{Eugene Agichtein}
\affiliation{Emory University}
\email{eugene.agichtein@emory.edu}

\begin{abstract}

An accurate understanding of a user's query intent can help improve the performance of downstream tasks such as query scoping and ranking. In the e-commerce domain, recent work in query understanding focuses on the query to product-category mapping. But, a small yet significant percentage of queries (in our website 1.5\% or 33M queries in 2019) have non-commercial intent associated with them. These intents are usually associated with non-commercial information seeking needs such as discounts, store hours, installation guides, etc. In this paper, we introduce Joint Query Intent Understanding (JointMap), a deep learning model to simultaneously learn two different high-level user intent tasks: 1) identifying a query's commercial vs. non-commercial intent, and 2) associating a set of relevant product categories in taxonomy to a product query. JointMap model works by leveraging the transfer bias that exists between these two related tasks through a joint-learning process. As curating a labeled data set for these tasks can be expensive and time-consuming, we propose a distant supervision approach in conjunction with an active learning model to generate high-quality training data sets. To demonstrate the effectiveness of JointMap, we use search queries collected from a large commercial website. Our results show that JointMap significantly improves both ``commercial vs. non-commercial'' intent prediction and product category mapping by 2.3\% and 10\% on average over state-of-the-art deep learning methods. Our findings suggest a promising direction to model the intent hierarchies in an e-commerce search engine.

\end{abstract}

\begin{titlepage}
\maketitle
\end{titlepage}

\section{INTRODUCTION AND RELATED WORK}

Query intent understanding is a key step in designing advanced retrieval systems like e-commerce search engines \cite{croft2010query}. Various approaches have been proposed to address query understanding such as 1) considering predefined high-level categories (i.e., informational, navigational, and transactional), 2) deploying semi-supervised learning with click graphs, 3) considering temporal query intent modeling, 4) understanding word-level user intent, and 5) applying relevance feedback and user behaviors. Although there has been a significant improvement in user intent inference, query understanding remains a major challenge \cite{zhang2019generic}.

E-commerce search queries have multiple intents associated with them. Ashkan et al. \cite{ashkan2009classifying} categorized search queries for e-commerce websites into commercial and nonc-commercial intents. However, Zhao et al. \cite{zhao2019dynamic} ignore the non-commercial queries due to small percentage of the search traffic. Commercial queries are queries with purchasing intent, while non-commercial queries cover a wide range of customer services (e.g., ``military discounts'' and ``installation guides'') as shown in Table. \ref{tab:Features}. 

\begin{table}
\footnotesize
\centering
    \begin{tabular}{l|l|l}
    \hline
    {\bf Search Queries}  & {\bf intent}& \bf Product Categories\\
    \hline \hline
    
     where is my shipped order& \textit{non-commercial} & \textit{-}\\
     how to install my tiles& \textit{non-commercial} & \textit{-}\\
     cost to rent a carpet cleaner& \textit{non-commercial} & \textit{-}\\
     18 volt ryobi& \textit{commercial} & [\textit{tools, electrical, lighting}]\\
     24 in. classic Samsung refrigerator & \textit{commercial} & [\textit{appliance, electrical}]\\
    \hline
    \end{tabular}
\caption{Dataset sample queries and their associated labels.}
\label{tab:Features}
\vspace{-0.8cm}

\end {table}

Query understanding in e-commerce search is challenging: 1) queries are often short, vague, and suffer from the lack of textual evidence \cite{ha2016large}, 2) small variation in textual evidence causes a drastic change in query intent; for example, ``30 in. 5.8 cu. ft. gas range installation kit'' has commercial intent but  ``30 in. 5.8 cu. ft. gas range installation'', has non-commercial intent, 3) product category mapping is a multi-label and non-exclusive problem. A practical solution must include a broader possible set of correct categories, while simultaneously keeping precision as high as possible \cite{zhao2019dynamic}, 4) there is class imbalance in both commercial vs. non-commercial and product category mapping tasks, because only a small fraction of data (1.5\% in our domain) has a non-commercial intent, and within the commercial queries, some product categories contain significantly more samples compared to others, and 5) commercial queries are easy to identify using user behavior information like click rates; however, that is not the case for non-commercial queries.

To address these problems, we introduce a new method of jointly learning query intent and category mapping, which allows transferring the inductive bias between these two relevant tasks. Also, we leverage label representation, which provides a richer representation to model the product categories. Finally, we propose an active learning algorithm to generate data for commercial vs. non-commercial intent. To address the class imbalance problem, we deploy focal loss, which is borrowed from computer vision.

Joint learning has been proposed as a practical approach to simultaneously learn relevant tasks due to the transfer of the inductive bias among them. Joint-learning finds applications in computer vision and natural language understanding \cite{khatri2018contextual}. Joint-learning improves the regularization and generalization of the learning models by utilizing the domain information \cite{caruana1997multitask}. In addition, with a joint model that addresses multiple tasks, only one model needs to be deployed; this contributes to reducing overhead and facilitates the maintenance of the system \cite{wang2018multi}. In this paper, we propose a joint-learning model that simultaneously learns both commercial and non-commercial query intents, and maps the incoming commercial query to a set of relevant product categories.

In this paper, we introduce a data-driven approach, which we call Joint Query Intent Mapping (JointMap). JointMap leverages the label representation proposed by Guoyin et al. \cite{wang2018joint} and modifies it to be applicable for a joint-learning task. JointMap also utilizes self-attention mechanism to improve the quality of the joint word-label attention vectors. For product category mapping, JointMap handles the imbalanced class problem using focal loss \cite{lin2017focal} which has been well-studied in the computer vision field to control the sparse set of candidate object locations. Finally, we propose an approach based on distant supervision in combination of active learning to generate both commercial and non-commercial queries.

In summary, our contributions are: 1) proposing a deep learning model to jointly learn product category mapping as well as users' non-commercial intents, 2) developing an active learning algorithm in conjunction with distant supervision to generate a user intent dataset from e-commerce data logs, and 3) modifying the joint word-category representation for query intent mapping tasks in e-commerce, as described in detail next. 

\vspace{-0.4cm}
\section{MODEL OVERVIEW}

\begin{figure}
\centering
\includegraphics[angle=-90,width = 70mm]{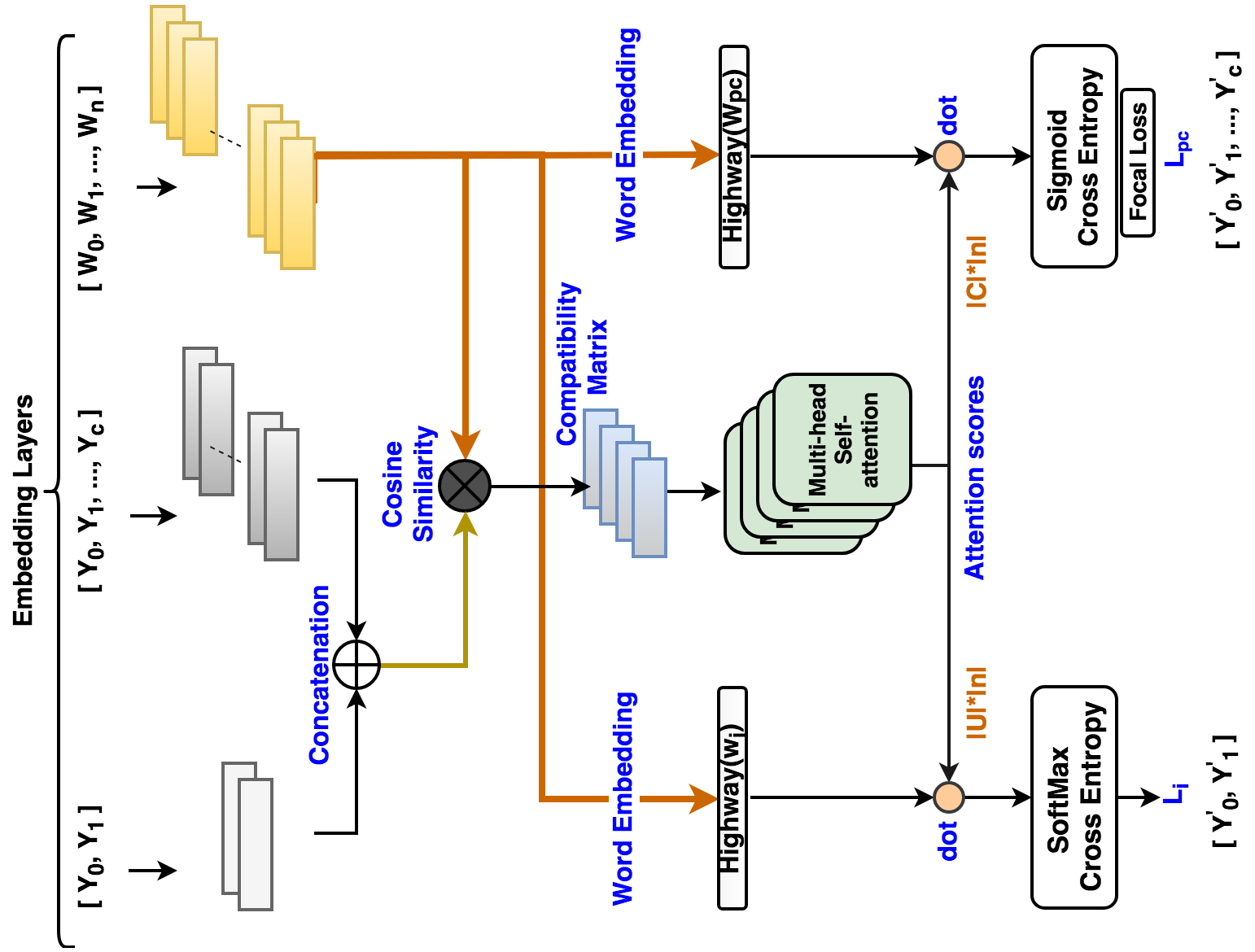}
\caption{ JointMap network architecture.}
\label{fig:joint-learning}
\vspace{-0.5cm}
\end{figure}

In this section, we present the network architecture of JointMap, as shown in Figure \ref{fig:joint-learning}. 
JointMap utilizes both word and category embeddings in which both representations are jointly trained to achieve an efficient semantic representation for a query. The proposed model consists of two deep learning layers: the first layer for the understanding of the user's commercial intent and the second layer for the prediction of relevant product categories in the taxonomy. As a result, the proposed model contains three embedding layers: a word embedding layer and two category embeddings layers, i.e., commercial vs. non-commercial and product-categories. Both category embedding types are concatenated, to compute the final product category representations. Then, a Compatibility Matrix (CM) is generated by computing the cosine similarity between the label and word representations. CM represents the relative spatial information among consecutive words (phrases) with their associated product category and commercial vs. non-commercial labels. Finally, CM is passed through a Multi-head self-attention layer to calculate attention scores. The word vectors simultaneously go through two Highway layers, and the output of each Highway is multiplied by their corresponding attention scores to generate the final query representation. Finally, the loss value of \begin{math}\mathcal{L}_{pc} \end{math} is computed using sigmoid cross-entropy for the product category mapping. Also, the loss value $\mathcal {L}_{i}$ is calculated using Softmax cross-entropy for determining the query's commercial intent.

In the next section we explain the details of the proposed model.

\vspace{-0.15cm}
\subsection{Joint-Learning of High-Level Intent Tasks}

We now introduce JointMap, a joint-learning model for high-level user intent prediction.

Suppose there is a search query dataset $ D =  \lbrace Q, C, U \rbrace$, where $Q$ is a set of search queries, $U$ represents user commercial vs. non-commercial intent, and $C$ is the candidate product category set. Each query consists of a sequence of words $q = [w_1; w_2; ... \ ;w_n]$ of size $n = 10$, and represents as $\mathbf{W}^{|W| \times V}$. Also, $C$ and $U$ are mapped to the embedding spaces $\mathbf{C}^{|C| \times V}$ and $\mathbf{U}^{|U| \times V}$, respectively. Then, the matrices $\mathbf{C}$ and $\mathbf{U}$ are concatenated to illustrate the whole label space. The word and label embeddings are initialized with Word2Vec and random embeddings of size $|V| = 300$, respectively. Cosine similarity between $\mathbf{L}$ and $\mathbf{W}$ is computed for each query $q$ to extract the relative spatial information among consecutive words with their associated labels, where $\otimes$ indicates the cosine similarity function.

\vspace{-0.3cm}
\begin{equation}
    \mathbf{H} = {(\mathbf{C} + \mathbf{U})}^{(|C|+|U|) \times V} \otimes \big (\mathbf {W}^{n \times V} \big )^T  
    \label{cosine}
\end{equation}
To extract the contribution of the words concerning their category, a multi-head self-attention mechanism with $n$ different heads is implemented on $\mathbf{H}$. Multi-head self-attention contains a parallel of linear projections of a single scaled dot-product function. Eq. \ref{head} shows a single head of the self-attention mechanism.

\vspace{-0.3cm}
\begin{equation}
        \mathbf{G} = SoftMax ( \frac{ \mathbf{H} K^T}{\sqrt{d_k}}) \mathrel{V}
        \label{head}
\end{equation}

where $K$ is the key matrix, $V$ is the value matrix, and $d_k$ is the dimension of the keys. Also, each projection is responsible for extracting the attention between word-label in a query and computes using weighted sum of the values. Next, $\mathbf{G}$ is split into two matrices of size $\mathbf{\hat{G}} = (|C| \times n) $ and $ \mathbf{\widehat{G}} = (|U| \times n) $. For both tasks, the word embedding vectors \textbf{W} are fed into a highway encoder layer, which has shown its effectiveness in improving network capacity \cite{zhang2019generic}. Then, the output is multiplied by their corresponding attention scores of $\mathbf{\hat{G}}$. 

\vspace{-0.35cm}
\begin{equation}
\mathbf{\alpha_1} =  Highway_1( \mathbf{W}),  \mathbf{\alpha_2} =  Highway_2( \mathbf{W})
\end{equation}
\vspace{-0.35cm}
\begin{equation}
\mathbf{\alpha_i} = sigmoid(r_w) \rightarrow r_w = relu(word2vec(w))
\end{equation}
\vspace{-0.35cm}
\begin{equation}
\mathbf{W_1} = \sum_{i=1}^{n}{\mathbf{\hat{G}}_i \times {\alpha}_1}
,
\mathbf{W_2} = \sum_{i=1}^{n}{\mathbf{\widehat{G}}_i \times {\alpha}_2}
\end{equation}
\vspace{-0.3cm}

Then, resulted $\mathbf{W_1}$ and $\mathbf{W_2}$ have the size of $(n \times V)$. They go through a fully connected layer to generate the semantic representations of both tasks. For product category mapping, a sigmoid cross-entropy loss function $ \mathcal{L}_{pc}$ is used since in sigmoid, the loss computed for every output $s_i$ is not affected by other component values. Also, a binary softmax cross-entropy loss $\mathcal{L}_{i}$ is applied to train the user commercial vs. non-commercial intent. 

\vspace{-0.4cm}
\begin{gather}
        \mathcal{L}_{pc} = - \sum_{c=1}^{|C|}{t_c \log \left ( {Sigmoid(s_c)}  \right ) } \\
     \mathcal{L}_{i} = - t_1 \log (SoftMax(s_1)) - (1-t_1)  \log (1- SoftMax(s_1))
     \vspace{-0.2cm}
\end{gather}

Where $s_c$ represents the prediction distribution and $t_c$ indicates the target labels. To address the class imbalance problem, particularly in the product category dataset, we update the loss values based on focal loss proposed in \cite{lin2017focal}. The focal loss was initially proposed for object detection and removing the effect of extreme foreground-background class imbalance in the images.  

\vspace{-0.4cm}
\begin{equation}
\mathcal{L}_{focal_{pc}} = \sum_{c=1}^{|C|}{\alpha_c \big( {Sigmoid(s_c)} - t_c \big)^\gamma \log \big({{Sigmoid(s_c)}}\big)}
\label{loss-cat}
\end{equation}

where $t$ is the target vector, $c$ is the class index, and $(f(s)- t)^\gamma$ is a factor to decrease the influence of well-classified samples.

\vspace{-0.1cm}
\subsubsection*{\bf JointMap overall loss: }

The final loss function is computed using a weighted loss over commercial vs. non-commercial, product category mapping intents.  

\vspace{-0.4cm}
\begin{equation}
\mathcal{L}_{total} = \beta_1 \mathcal{L}_{focal_{pc}} + \beta_2 \mathcal{L}_{i}
\label{overall-loss}
\end{equation}

\section{DATASET OVERVIEW}
\label{sec:Dataset}

In this section, we describe the dataset collected from search logs of a large e-commerce search engine in July 2019, and provide details the algorithms used for generating user-intent datasets. We propose an algorithm to simultaneously generate both datasets, which consists of three steps: 1) generating the commercial vs. non-commercial queries, 2) oversampling of the non-commercial queries to balance the dataset, and 3) creating the product category dataset based on the commercial queries. Algorithm. \ref{Alg-cvsnc} represents the steps for generating commercial vs. non-commercial samples. In this method, first we need to generate a small-size dataset that covers all expected non-commercial intents (e.g., ``installation guides'').

\begin{algorithm}
\SetAlgoLined
\KwResult{\textit {Commercial Vs. Non-commercial Dataset}}
 D\_init = A small-size Dataset by human supervision \;
 Test = Hold-out test set\;
 \While{ Accuracy < threshold}{
      \textbf{D} = Expand(D\_init) using KNN;\\
      Confidence Scores = SVM(\textbf{D});\\
      TS = Find(tricky samples) using confidence scores;\\
      \textbf{D} = Re-label(TS) using human supervision;\\
      D\_init = \textbf{D};\\
      Accuracy = Compute\_Accuracy(Test);\\
 }
 \caption{Commercial vs. non-commercial dataset.}
 \label{Alg-cvsnc}
\end{algorithm}

Then, we over-sample the non-commercial queries as described in \cite{charte2019dealing} to make the dataset balanced (only 1.5\% of the queries have a non-commercial intent). Similar to \cite{zhao2019dynamic}, we utilize user behavior data like click rate, to generate the category labels associated with each commercial query. Algorithm. \ref{Alg-pd} describes different steps to create the product mapping dataset.

\vspace{-0.2cm}
\begin{algorithm}[h]
\SetAlgoLined
\KwResult{\textit {Product Category Dataset}}
Product\_category = \{\};\\
  \For{each query in Q}{
  pid\_list =  Extract(pid that user clicks)\\
  \For{pid in pid\_list}{
      category\_list= Find(category(pid) in taxonomy)
      }
  \For{category in  category\_list}{
      \If {if click\_rate > r}
      {
          product\_category(query).add(category)
      }
  }
 }
\caption{Product category dataset generator.}
\label{Alg-pd}
\end{algorithm}
\vspace{-0.2cm}
Finally, a dataset of size 195K with 32 product categories such as \textit{tools, appliance, outdoors, etc.} extracted from the search logs.

\vspace{-0.1cm}
\section{EXPERIMENTAL SETUP}

In this section, we describe the parameter setting, metrics, baseline models, and experimental procedures used to evaluate JointMap.

\vspace{-0.1cm}
\subsubsection*{\bf Parameter Setting. }
We used Adam optimizer with a learning rate of $\eta=0.001$ and a mini-batch of size 64 for training. The dropout rate of 0.5 is applied at the fully-connected and ReLU layers to prevent the model from overfitting. 

\vspace{-0.1cm}
\subsubsection*{\bf Evaluation Metrics. }

To evalate JointMap, both Micro- and Macro- averaged F1-score for both tasks are reported.
\vspace{-0.1cm}
\subsubsection*{\bf Methods Compared. }
\label{sec:comp}
We summarize the \textbf{multi-label} classification methods compared in the experimental results.
\begin{itemize}
    \item {\bf Tf*idf + SVM:} One-Vs-Rest SVM with a linear kernel.
    \item {\bf VDCNN:} Very Deep Convolutional Neural Network \cite{conneau2016very}.
    \item {\bf FastText:} Text classification method developed by Facebook\cite{bojanowski2017enriching}.
    \item {\bf LEAM:} Word-label representation model\cite{wang2018joint}.
    \item {\bf XML-CNN:} Extreme multi-label text classification \cite{liu2017deep}.
    \item {\bf JointMap:} The proposed model.
\end{itemize}

\subsubsection*{\bf Dataset Experimental Design.} 
\begin{figure}
\centering
\includegraphics[width = 70mm]{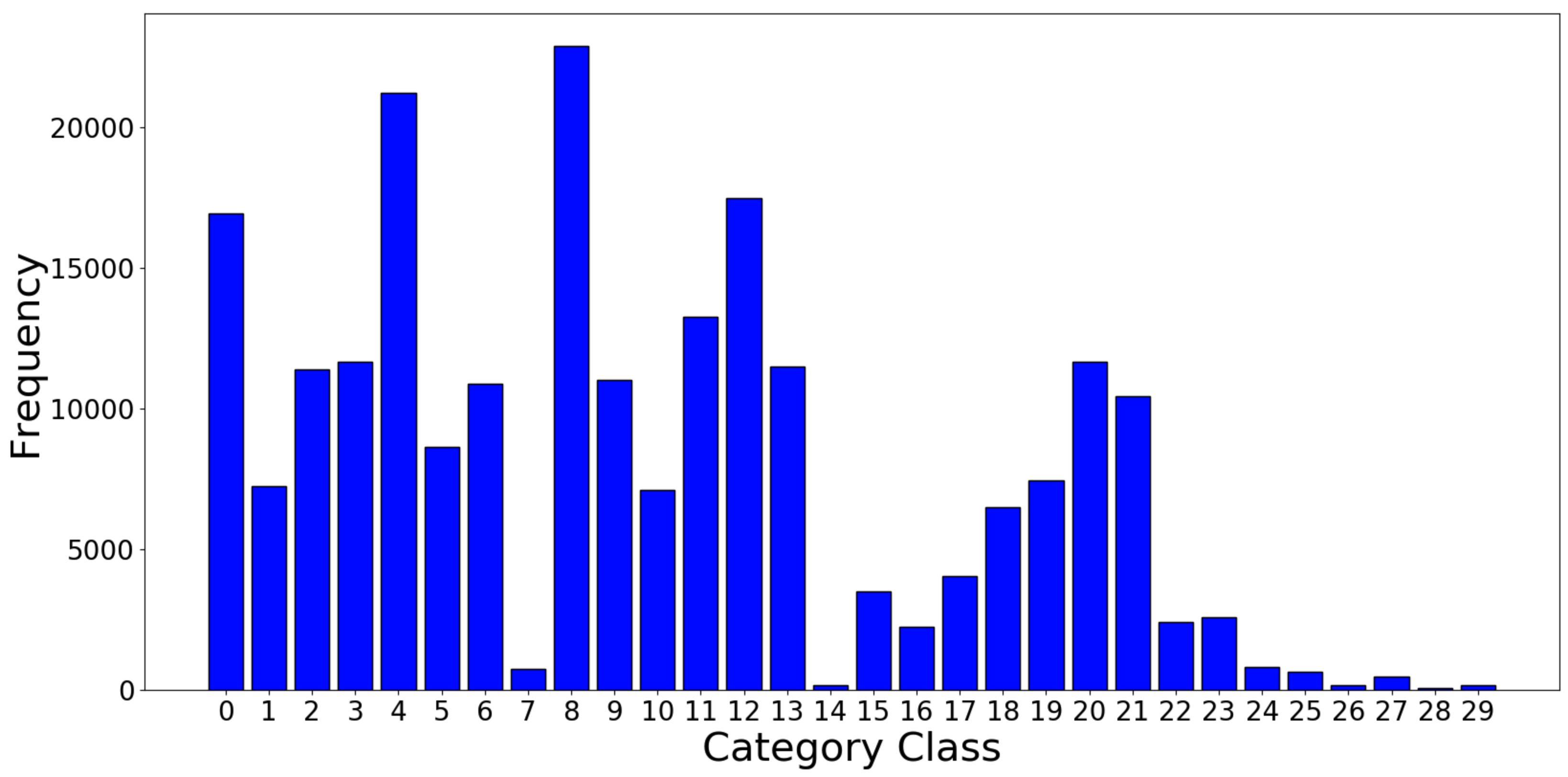}
\caption{Product category distribution.}
\label{fig:dist}
\vspace{-0.5cm}
\end{figure}

We use an SVM model with n-gram tf*idf as features to perform distant supervision method due to multiple reasons: 1) SVM is fast and scalable, 2) the features and results are interpretable for supervisors, 3) SVM has proved its effectiveness on text data, 4) SVM provides confidence scores to detect the tricky samples. Moreover, two different human annotators were asked to label 540 samples manually. The (Matching, Kappa) scores of (0.98, 0.96) are computed, which is a “significant agreement.” The category distribution is shown in Figure \ref{fig:dist}. 

\vspace{-0.2cm}
\subsection{Main Results and Ablation Analysis}

To evaluate the models described in Section \ref{sec:comp}, 70\% of the dataset is used for training, 10\% for validation, and 20\% for test. Table \ref{tab:overall} summarizes the performance of the models. The results are reported for both commercial vs. non-commercial classification and product category mapping. All the improvements are statistically significant using a one-tailed Student's t-test with a p-value < 0.05.

\vspace{-0.3cm}
\begin{table}[h]
\footnotesize
\centering

\begin{tabular}{@{}l||ll||ll@{}}
\multicolumn{5}{c}{\hspace{45pt} \bf{Dataset}} \\ 
\multicolumn{1}{c||}{\textbf{Method}}&
\multicolumn{2}{c}{Commercial vs. Non-commercial} & 
\multicolumn{2}{c}{Product Category Mapping} \\
\cline{2-5}
\multicolumn{1}{c||}{}  & \multicolumn{1}{c}{Macro-F1} & \multicolumn{1}{c||}{Micro-F1} & \multicolumn{1}{c} {Macro-F1} & \multicolumn{1}{c}{Micro-F1} \\
\bottomrule\bottomrule
tf*idf+SVM   & 90.71    &  90.26   & 48.75  & 76.84 \\
\hline

VDCNN \citep{conneau2016very} &  91.28 & 91.34 &  51.41    & 79.34 \\
FastText \citep{bojanowski2017enriching}& 92.18 & 92.15  & 60.06 & 79.69 \\
XML-CNN \citep{liu2017deep} &  93.11  & 93.01  &  58.40     & 81.61 \\
LEAM \citep{wang2018joint} &  93.96  & 93.66  &  58.90     & 81.31 \\
\hline
\hline
JointMap      &  94.80 \bf(+1.1\%)  & 94.63 \bf (+1.0\%)  &  62.60 \bf (+6.3\%) & 83.01 \bf (2.1\%)    \\

\bottomrule
\end{tabular}
\caption{Macro- and Micro- averaged F1 for different models. The improvements reported against LEAM.}
\label{tab:overall}
 \vspace{-0.7cm}
\end{table}

For the user commercial intent mapping task, the results indicate that the Macro-averaged F1 improves 4.5\%,3.8\%,2.8\%,1.0\%, and 1.8\% compared to tf*idf, VDCNN, FastText, LEAM, and XML-CNN models respectively. In product category mapping task, the improvements are more significant. There is improvement of 28.4\%, 22.1\%, 4.2\%, 6.3\%, and 7.2\% over tf*idf, VDCNN, FastText, LEAM, and XML-CNN models, respectively. As a results, JointMap improves macro-averaged F1 scores over current state-of-the-art deep learning models by 2.3\% on commercial vs. non-commercial intents, and a 10\% improvement over product category mapping.

In reference to user commercial intent prediction, a 2.3\% improvement is considerable since it is in the context of a large e-commerce search engine that receives billions of search queries per year. For product category mapping, the F1-averaged macro experiences a higher jump when compared to the F1-averaged micro (6.3\% vs. 2.1\%). This improvement indicates the positive impact of inductive bias between these two tasks, which not only boosts the performance of majority classes, but it also contributes to minority classes. For instance, the Macro-average F1 for 8-button minority classes shows in Figure. \ref{fig:dist} for XML-CNN and LEAM are 21.76\% and 18.33\%, respectively, while this number jumps to 31.28\% for JointMap.

\vspace{-0.15cm}
\subsubsection*{\bf Focal Loss Impact.} Using focal loss deteriorates the overall micro- and macro- averaged F1-scores by 0.6\%, 1.5\%, respectively. However, the macro-average F1 on 8-button minority classes without focal loss is 31.28\%, while with presence of focal loss is 33.81\%. This shows a relevant improvement of 8.1\%. Also, we observe that in absence of focal loss, the performance of at least two of the minority classes is 0\%, therefore making the use of focal loss necessary.

\vspace{-0.15cm}
\subsubsection*{\bf Parameter Tuning.} To evaluate the impact of hyper-parameter tuning in JointMap, we implemented a grid search approach on $\beta_1$ and $\beta_2$ in Eq. \ref{overall-loss}. We observed that using a smaller $\beta$ for each task causes a slower convergence for that specific task. However, the final results is not significantly different. In our experiments, a simple average works as good as a fine-tuned hyper-parameter model. For focal loss hyper-parameter tuning, we repeat the experiments with different $\gamma$ values of 1, 1.2, 1.5, and 2. We observed that the best results achieve using the $\gamma =1.5$, where the original paper suggested using $\gamma = 2$ for computer vision application.

\vspace{-0.1cm}
\section{CONCLUSIONS AND FUTURE WORK}
We introduced JointMap, a deep learning model designed for jointly learning two high-level intent tasks on e-commerce search data. JointMap utilized word and label representations and leveraged focal loss to tackle class imbalance problem in catalog categories. Our results were promising compared to the state-of-the-art deep learning models with an average raise of 2.3\% and 10.9\% on Macro-averaged F1 in user commercial vs. non-commercial intent and product category mapping, respectively. Our future work includes tuning the JointMap model incorporate contextual information within a session. In summary, the presented work advances the state-of-the-art user intent prediction, and lays the groundwork for future research on user intent understanding in e-commerce. 
\vspace{-0.1cm}
\subsubsection*{\bf Acknowledgements}
We gratefully acknowledge the financial and computing support from The Home Depot Search \& NLP team.
\vspace{-0.2cm}
\bibliographystyle{abbrv}
\bibliography{REFERENCES}
\balance

\end{document}